\documentclass[11pt]{llncs}
\usepackage{graphicx}
\usepackage{cite}
\usepackage{url}
\usepackage{times}
\usepackage{mathfont}

\def\sin{\mathop{\rm sin}}
\def\cos{\mathop{\rm cos}}

\newtheorem{open}{Open Problem}

\DeclareSymbolFont{AMSb}{U}{msb}{m}{n}
\DeclareSymbolFontAlphabet{\Bbb}{AMSb}
\def\Real{\ensuremath{\Bbb R}}

\mathcode`O="724F

\makeatletter
\def\hb@xt@{\hbox to }
\makeatother

\let\oldendproof\endproof
\def\endproof{\qed\oldendproof}

\newcommand{\OCfigure}[3]{
\begin{figure}[t]
 \centering\includegraphics[width=#3]{#1}
 \caption{#2}
 \label{#1}
\end{figure}}

\let\SCfigure\OCfigure

\begin{document}

\title{Flipping Cubical Meshes} 

\author{Marshall Bern\inst{1},
David Eppstein\inst{2}, and
Jeff Erickson\inst{3}}

\institute{Xerox Palo Alto Research Center\\
3333 Coyote Hill Rd., Palo Alto, CA 94304\\
\email{bern@parc.xerox.com}
\medskip
\and
University of California, Irvine\\
Dept. of Information \& Computer Science,
Irvine, CA 92697\\
\email{eppstein@ics.uci.edu}
\medskip
\and
Univ. of Illinois, Urbana-Champaign\\
Dept. of Computer Science,
1304 W. Springfield Ave.,
Urbana, IL 61801\\
\email{jeffe@cs.uiuc.edu}}

\maketitle   

\begin{abstract}We define and examine flip operations
for quadrilateral and hexahedral meshes, similar to the flipping
transformations previously used in triangular and tetrahedral mesh
generation.

\medskip
{\bf Keywords:} quadrilateral meshing, hexahedral meshing, flip graph
connectivity, mesh improvement.
\end{abstract}

\section{Introduction}

In this paper we propose flipping transformations for cubical meshes,
such as quad and hex meshes. 
These flipping transformations are precise analogues of the 
well-known flipping transformations for simplicial meshes.  
We did not invent these transformations. 
They have already appeared---although they were not 
explicitly enumerated---in the mathematical
polytope literature~\cite{BilStu-AM-92,Zie-95}. 
We have not found any previous references 
in the meshing literature. The closest work seems
to be the quad refinement operations of Canann,
Muthukrishnan, and Phillips~\cite{CanMutPhi-EC-98}
and the hex reconnection primitives of 
Knupp and Mitchell~\cite{KnuMit-TR-99}.

As in the simplicial case, quad and hex flipping could be useful for mesh
generation, improvement, and refinement/derefinement. We also use
flipping to study existence questions for hexahedral
meshes.  Up until this present paper, the theoretical results on
quad-to-hex extension have been positive, suggesting that  (under
simple necessary conditions) it should always be possible.
Mitchell~\cite{Mit-STACS-96} and Thurston~\cite{Thu-93} independently
showed that any quad surface mesh with an even number of quadrilaterals,
and topologically equivalent to a sphere, can be extended to a {\it
topological\/} hex volume mesh, which allows elements to have warped,
nonplanar sides.  (An even number of surface quadrilaterals is a
necessary condition, because hex elements each have six sides,  and each
interior face will account for two hex sides.) 
Eppstein~\cite{Epp-CGTA-99} gave another proof of this result and
improved the complexity bound from $O(n^2)$ to $O(n)$ elements. Each of
the two proofs generalize to certain more topologically  complicated
inputs as well.  However, it has not been clear under what circumstances
these topological meshes can be realized using flat-faced cuboids. In
this context we discuss a simple 10-element boundary complex, the {\em
bicuboid}, which we have been unable to mesh, and use flipping to show
that many possible topological meshes for this complex are unrealizable.

This is primarily a theoretical paper---realizability
is not of great practical importance since the
isoparametric trilinear elements typically used with unstructured
hexahedral meshes allow the cells to have curved boundaries.
However, the existence or nonexistence of flat-faced meshes may
shed light on the question of how many curved elements are needed to achieve satisfactory mesh quality.  In addition,
our study of flip graph connectivity relates to mesh
improvement algorithms: by characterizing the connected components of the
quadrilateral flip graph for simply-connected regions
(Theorem~\ref{thm:qfc}) we show that any local improvement operation
that remeshes bounded regions of a quadrilateral mesh can be simulated
by a sequence of simpler flip and parity-change operations.

\section{Definitions}\label{def-sec}

A {\it polyhedron\/} is a set of closed planar
polygonal faces in $\Real^3$, that intersect
only at shared vertices and edges, and such that each edge is shared
by exactly two faces and each vertex is shared by exactly one
cycle of faces.\footnote{ This definition of polyhedron allows  
multiple connected components (two cubes can be one polyhedron!),
as well as voids and tunnels.
It disallows non-manifold boundaries, such as those used to 
model objects made from two different materials.}
A polyhedron divides $\Real^3$ into a bounded
{\it interior\/} and an unbounded {\it exterior\/}, each of which
may contain more than one connected component.

A {\it cuboid\/} in $\Real^d$ is a convex polytope, 
combinatorially equivalent to the unit cube $[0,1]^d$.  
{\it Combinatorial equivalence\/} means that the faces 
(vertices, edges, and so forth) of the cuboid are
in one-to-one correspondence with the faces of the unit cube,
and that this correspondence preserves intersections. 
In $\Real^2$ a cuboid is simply a convex quadrilateral,
called a {\it quad\/} for short. 
In $\Real^3$ a cuboid is a convex polyhedron called a {\it hex\/}. 

A {\it quad surface mesh\/} is a polyhedron
whose faces are all quads.
A {\it hex mesh\/} is a quad surface mesh, along with interior quads 
that subdivide the polyhedron interior into hexes.  
The intersection of any pair of hexes is either empty, a single
vertex, a single edge, or a single quad. 
Quad surface meshes and hex meshes are both examples
of {\it cubical complexes\/} \cite{BliBli-DM-98,JosZie-DCG-00,math.GT/0204085}, the analogue of simplicial complexes in which each $k$-dimensional face is a $k$-dimensional cuboid.

We also consider {\em topological hex meshes}: cubical complexes
homeomorphic to the input domain, with a boundary mesh combinatorially equivalent to that of the input.  In this context,
we call the quad and hex meshes defined earlier {\em geometric meshes}
to emphasize the geometric requirements on their cell shapes. Topological
meshes can be {\em realized} by specifying locations in the domain for
each internal vertex of the complex; the realization is a hex mesh if
each cell's eight vertex locations have a convex hull that is a cuboid,
and no two of these cuboids have intersecting interiors. A realization is
{\em self-intersecting} if each four vertices of a quadrilateral in the
mesh have locations that form a convex quadrilateral in the domain,
without respect for whether these quadrilaterals form the boundaries of
disjoint cuboids; a self-intersecting mesh can be viewed as a
generalized geometric mesh in which some cells are allowed to be
inverted.

\section{Flipping}\label{flipping-sec}

\OCfigure{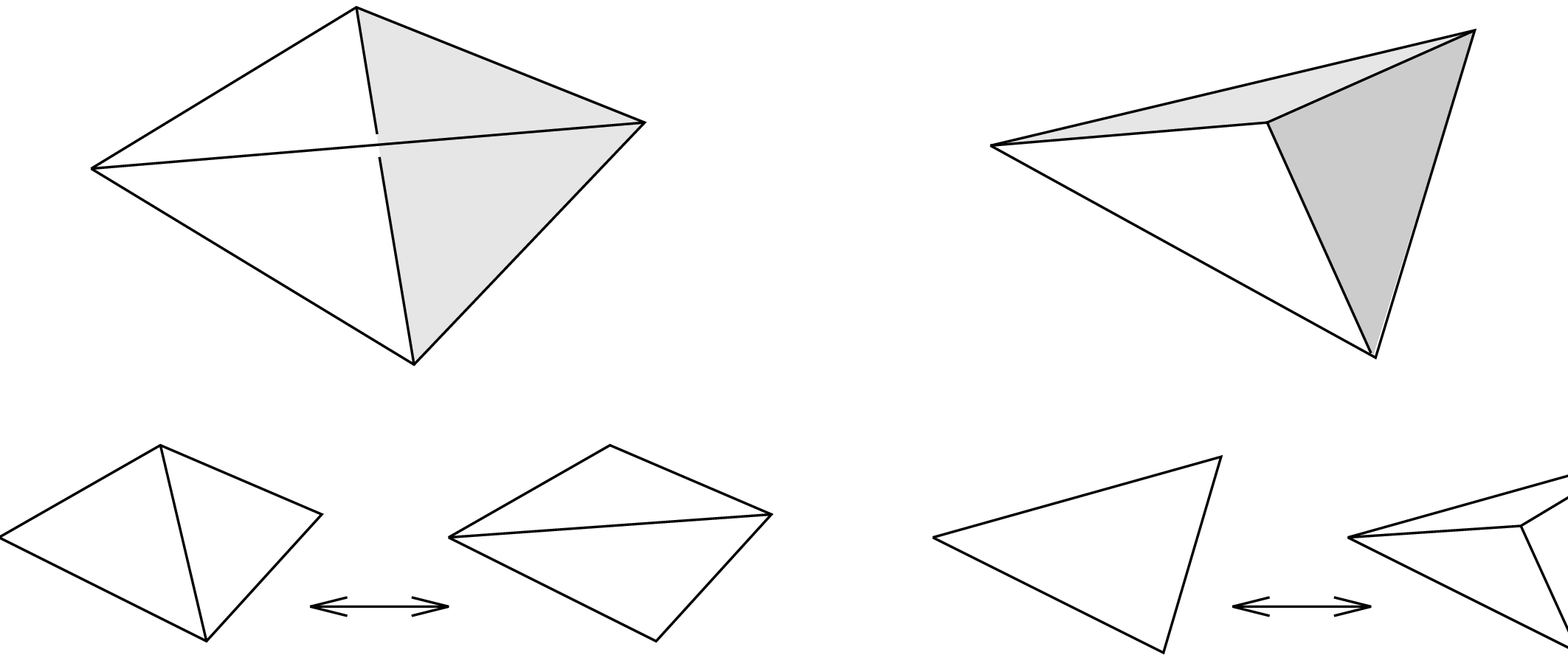}{The 2-2 and 1-3 flips for triangular meshes
correspond to switching upper and lower facets of a tetrahedron.}
{4.2in}

\OCfigure{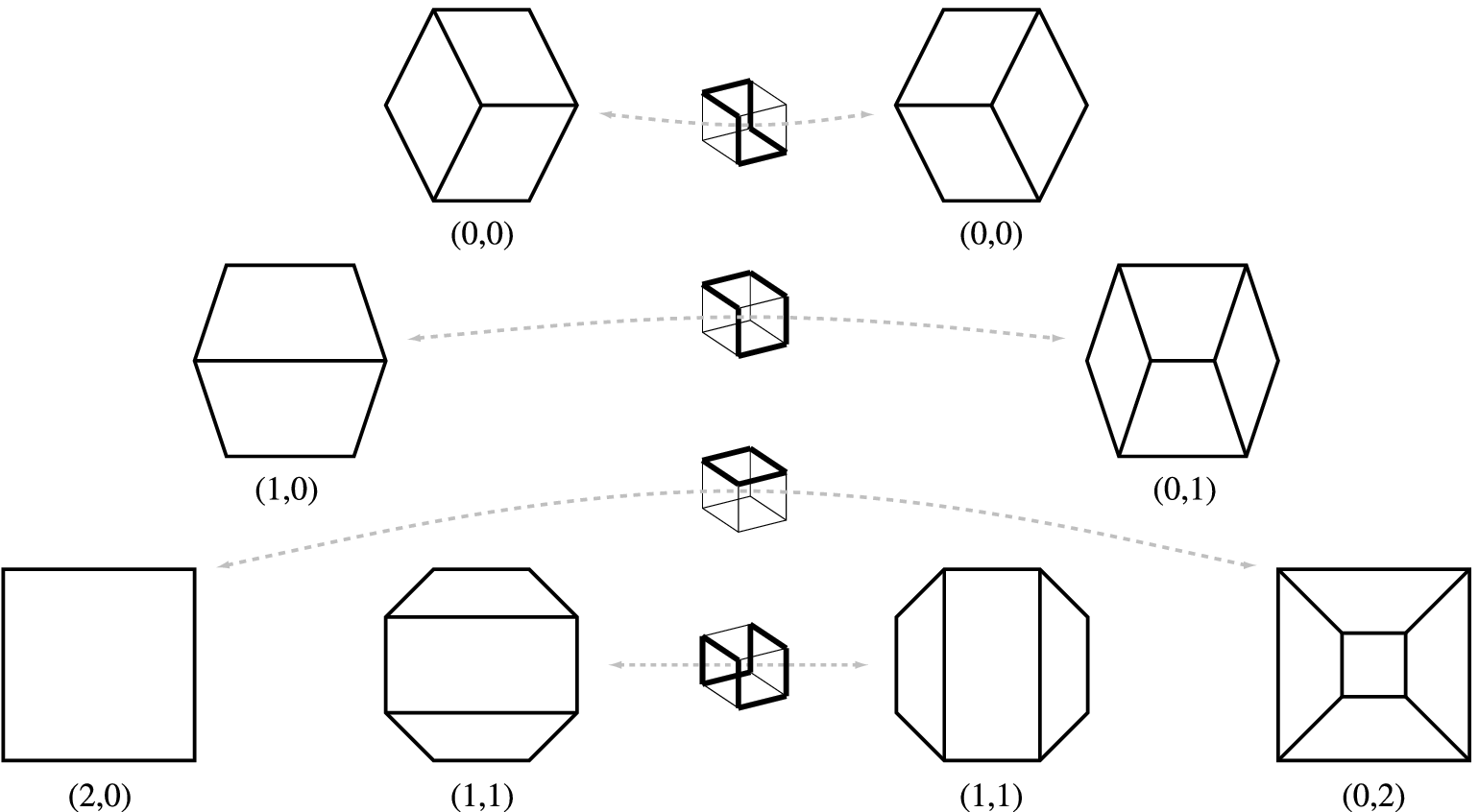}{Flips for quad
meshes correspond to switching upper and lower facets of a
three-dimensional cuboid.  The bold cuboid edges mark the
division between upper and lower facets.} {6.5in}

Consider the well-known {\it flip\/} transformation
for a triangular mesh, which
switches one diagonal for the other in a convex quadrilateral
$Q$ formed by two triangular elements sharing a side. 
We can think of this 2-2 flip (two elements in and
two elements out) as exchanging the upper and lower
facets of a three-dimensional simplex $P$ that projects to $Q$,
as shown on the left in Figure~\ref{triflips}.
This view leads us to consider the 
refinement transformation that subdivides a triangle into
three smaller ones by adding an interior vertex
as a 1-3 flip (Figure~\ref{triflips} right side). 
The reverse of the 1-3 flip is a 3-1 flip, which
derefines the triangulation by removing a vertex.

The almost-as-well-known flips for tetrahedral meshes
can be defined analogously by the exchange of the upper and lower
facets of a four-dimensional simplex.  
The flips include the 2-3 flip, which exchanges two tetrahedra
sharing a face for three sharing an edge, and the 1-4 flip
which subdivides a tetrahedron by adding an interior vertex,
along with their reverse flips.

This view of flipping extends to quad and hex meshes.
The flips for quad meshes are induced by the combinatorially
distinct exchanges of upper and lower facets of a three-dimensional cuboid,
as shown in Figure~\ref{jeffqf}, and the flips for hexahedral meshes are induced by the combinatorially distinct exchanges of upper and lower facets of a four-dimensional hypercube, as shown in Figure~\ref{hextriangle}.

The facets of a $d$-dimensional cube or hypercube can be grouped into $d$ opposite pairs.
In order to discover all possible flips, we describe any set of facets of a cube or hypercube by how it relates to these pairs.  To any set $S$ of cube or hypercube facets, we assign
a pair of integers $(X,Y)$, where $X$ denotes the number of pairs of facets eXcluded from $S$, and $Y$ denotes the number of pairs of facets Yncluded in $S$.
The remaining $d-X-Y$ pairs of facets have one of their two facets included in $S$, and the other facet excluded from $S$.  The symmetries of the hypercube include all permutations of opposite pairs of facets,
and all swaps of the two facets in each pair, so any two sets $S$ described by the same pair $(X,Y)$ are combinatorially equivalent.

The meshes in Figures \ref{jeffqf}
and~\ref{hextriangle} are labeled by these pairs of numbers.  If a set $S$ is represented by a pair $(X,Y)$, the complementary set is represented by pair $(Y,X)$.  So, a flip can be described as the replacement of a set of mesh cells combinatorially equivalent to the set $(X,Y)$ by a different set of cells combinatorially equivalent to $(Y,X)$.  In order to find all possible flips, it remains only to determine which pairs $(X,Y)$ form topological disks.

\begin{lemma}
Let $S$ be a set of facets of a $d$-hypercube described by pair $(X,Y)$.
Then $S$ forms a topological disk if and only if $X+Y<d$.
\end{lemma}

\begin{proof}
If $X+Y<d$, there is some facet $f\in S$ such that the opposite facet is not contained in $S$; then $f$ is adjacent to every facet in $S$.  We can retract all facets to $f$ by an affine transformation that shrinks the coordinate axis perpendicular to $f$; therefore $S$ has a retraction to a $(d-1)$-dimensional disk and is itself a disk.

On the other hand, suppose $X+Y=d$.  We show by induction on $X$ that $S$ has a retraction to a $Y$-sphere, and therefore cannot be a disk.  More specifically, we can retract $S$ in a facet-preserving way to the boundary of a $Y$-hypercube.
As a base case, if $X=0$ and $Y=d$, $S$ is the whole boundary of the $d$-hypercube.
Otherwise, $S$ is formed by removing a pair of facets from a set $S^+$ that can be retracted to a $(Y+1)$-hypercube $C^+$.  Removing the final two facets from $S^+$ corresponds, in its retraction, to removing two opposite facets from the boundary of $C^+$, and the remaining facets of $C^+$ can be contracted to a $Y$-hypercube by again using an affine map that contracts the coordinate axis perpendicular to the removed facets.
\end{proof}

We cannot allow sets $S$ that do not form disks to count as flips, because exchanging such a set for its complement would change the topology of the domain.
The possible flips in a $d$-dimensional cubical mesh correspond to subsets of a $(d+1)$-hypercube, and are therefore given by the pairs $(X,Y)$--$(Y,X)$ where $X+Y\le d$.

\begin{corollary}
The $d$-dimensional cubical meshes have
$(d+1)(d+2)/2$ combinatorially distinct flippable submeshes $(X,Y)$
which can be grouped into $\lfloor (d+2)^2/4 \rfloor$ flip pairs $(X,Y)$--$(Y,X)$.
\end{corollary}

The possible quadrilateral mesh flips are therefore
$(2,0)$--$(0,2)$ (changing one quadrilateral for five),
$(1,0)$--$(0,1)$ (changing two quadrilaterals for four),
and two changes of three for three quads:
$(0,0)$--$(0,0)$ and $(1,1)$--$(1,1)$.
along with their reverses. We can picture the $(2,0)$ mesh as corresponding to the unique lowest face of a cuboid, and the flipped $(0,2)$ mesh as the unique highest cuboid face together with the four neighboring side faces. The $(1,0)$--$(0,1)$ flip can similarly be pictured with
unique lowest and highest edges, and the $(0,0)$--$(0,0)$ flip can be pictured with unique lowest and highest vertices.  The remaining
$(1,1)$--$(1,1)$ flip transforms the top, front, and back faces of a cube into the bottom, left, and right faces.

\OCfigure{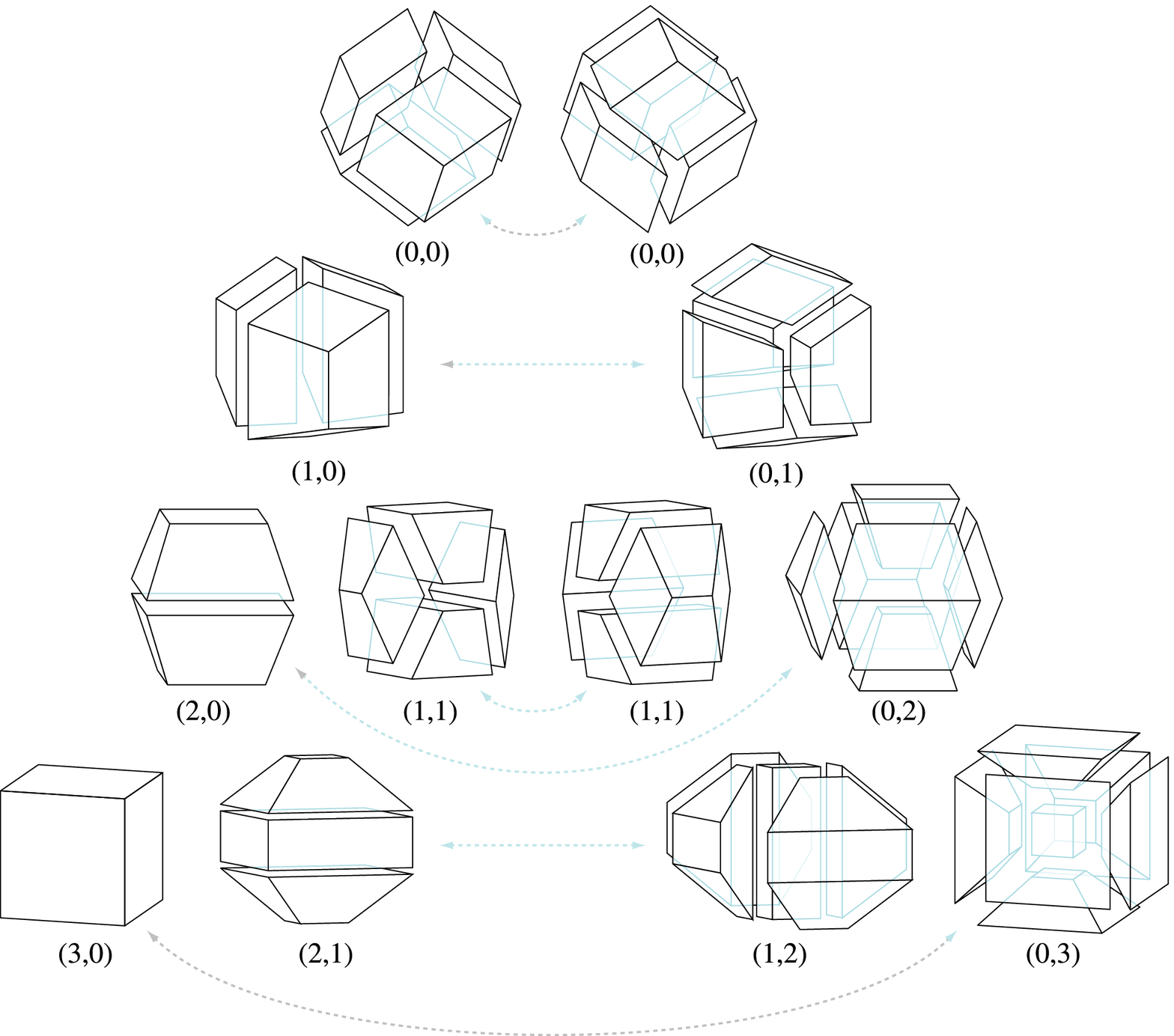}{The flips for hex
meshes correspond to switching upper and lower facets of a
four-dimensional cube.}{6.5in}

Analogously, the flips for hex meshes are induced by all possible
exchanges of upper and lower facets of a four-dimensional cuboid, as
shown in Figure~\ref{hextriangle}. The $(3,0)$--$(0,3)$ flip converts one hex to seven, and
can be pictured with unique lowest and highest facets of a hypercuboid in
$\Real^4$. The $(2,0)$--$(0,2)$ converts two hexes to six, and can be pictured with unique highest and lowest two-dimensional
faces of a hypercuboid. There are two different flips from three to five hexes: $(1,0)$--$(0,1)$ and
$(2,1)$--$(1,2)$.  In the $(2,1)$ three-hex mesh, the hexes are connected in a path. This mesh  can be pictured most symmetrically with a unique lowest hypercuboid facet
bordered by two other mutually disjoint almost-lowest facets, and the complementary $(1,2)$ mesh can be pictured with a unique highest facet (a sort of hex
tunnel in Figure~\ref{hextriangle}) surrounded by four other high facets.
The other three-hex mesh in our set of flips, $(1,0)$, has three hexes connected in a cycle,
and can be pictured  with a unique lowest hypercuboid
edge surrounded by three facets; its complementary five-hex mesh, $(0,1)$, has a
unique highest edge, whose endpoints do not appear on the lower convex
hull, surrounded by three facets.  Finally, there are two different flips involving sets of four hexes.
The $(1,1)$ flip corresponds to a lower-convex-hull facet connectivity of a clique minus an edge, and can be pictured with a unique lowest two-dimensional hypercuboid face; the $(0,0)$ flip has four mutually adjacent hexes, and can be pictured with a unique lowest hypercuboid vertex.

The regions of $\Real^3$ occupied by the cells of a flip are all combinatorially inequivalent:
a cuboid for the $(3,0)$--$(0,3)$ flip,
a bicuboid for the $(2,0)$--$(0,2)$ flip,
a twisted rhombic dodecahedron for the $(1,0)$--$(0,1)$ flip,
a tricuboid for the $(2,1)$--$(1,2)$ flip,
a fourteen-faced shape resembling a square orthobicupola for the $(1,1)$--$(1,1)$ flip, and a rhombic dodecahedron for the $(0,0)$--$(0,0)$ flip.
These shapes can be determined from the two-dimensional quadrilateral flips:
if one forms a polyhedron from the convex hull of two parallel copies of
an $(X,Y)$ quadrilateral mesh, the result is the region of an $(X,Y+1)$--$(Y+1,X)$ flip
(for instance, the $(0,2)$ bicuboid is formed by the convex hull of two parallel two-quadrilateral $(0,1)$ meshes), and if one instead forms the convex hull of an $(X,Y)$ mesh and a parallel $(Y,X)$ mesh, the resulting three-dimensional shape is the region for an $(X,Y)$--$(Y,X)$ flip.

Three of the hexahedral flips are listed by Knupp and
Mitchell~\cite{KnuMit-TR-99}: the $(3,0)$--$(0,3)$ flip is given as an example of
their ``pillowing'' transformation, the $(2,0)$--$(0,2)$ flip is given as an example
of their ``inflating hex ring'' transformation, and the $(0,0)$--$(0,0)$ flip
is called the ``rotate three hexes primitive'' despite its symmetric
action on four hexes.  The $(3,0)$--$(0,3)$ and $(2,0)$--$(0,2)$ flips are also used by
Mar\'echal~\cite{Mar-IMR-01} to reduce the complexity of some case
analysis in his conformal mesh refinement algorithm.

\section{Flippability}

\OCfigure{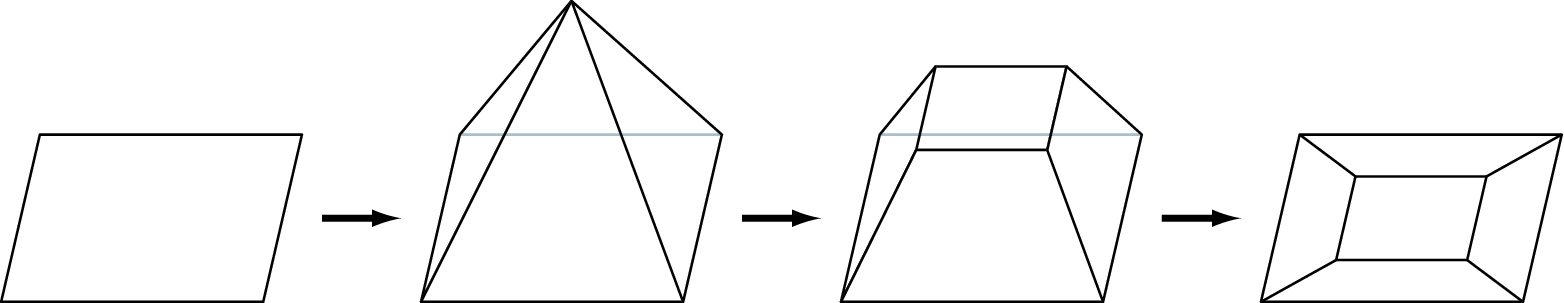}{The $(2,0)$--$(0,2)$ quad flip and $(3,0)$--$(0,3)$ hex flip can be performed
geometrically by projecting the truncation of a higher dimensional
pyramid.}{5.5in}

We say that a set of cells in a mesh is {\em flippable} if it can be
replaced by one of the flips described above.  For topological meshes,
a set of cells is flippable whenever it has the appropriate connectivity
pattern, but for geometric meshes additional conditions will likely
be required.  For instance, in triangular meshes, a pair of triangles is
flippable if and only if it forms a convex quadrilateral.

We say that a hex mesh is {\em generic} if no three vertices are
collinear and no two quadrilaterals are coplanar. We say that a type of flip is {\em automatically flippable}
if, whenever the configuration of cells on one side of the flip occurs
in a generic geometric or self-intersecting mesh, the flip can be
performed resulting in another generic (possibly self-intersecting)
mesh.  That is, for these types of flips, flippability of geometric
meshes may involve convexity constraints disallowing inverted elements,
but does not involve additional constraints on the flatness of cell
boundaries. In triangular, quadrilateral, or tetrahedral
meshes, all flips are automatically flippable, because all
cell boundaries are always flat.  The requirement that each
quadrilateral face be flat leads to more interesting theory in the
hexahedral case.

\begin{theorem}\label{automatic}
The $(3,0)$--$(0,3)$, $(0,3)$--$(3,0)$, $(2,0)$--$(0,2)$, $(0,2)$--$(2,0)$, and $(0,0)$--$(0,0)$ hex mesh flips are automatically flippable.
\end{theorem}

\begin{proof}
The result is trivial for the $(0,3)$--$(3,0)$ flip, since it just involves removal
of some faces from the $(0,3)$ mesh.  We can flip the $(3,0)$ mesh by viewing the
initial cuboid as the base of a pyramid in four dimensions, forming a
four-dimensional cuboid by truncating the pyramid's apex by a
hyperplane, and replacing the base by the orthogonal
projection of the truncated pyramid's remaining faces
(Figure~\ref{frustum}).  There are enough degrees of freedom in this
construction to preserve genericity.

To understand the $(2,0)$--$(0,2)$ flips, let's analyze the conditions under
which the six-cuboid $(0,2)$ mesh is possible. This mesh (shown in the upper
right corner of Figure~\ref{hextriangle}) breaks a {\em
bicuboid} into a {\em ring} of four cuboids connected to the bicuboid
sides, and a {\em core} of two cuboids connected to the top and bottom.

Let's first look at the ring. The planes $P_i$ ($i=0,1,2,3$) where its
cuboids connect to each other are determined by triples of
points on the bicuboid.  Each such plane contains four ring vertices,
three of which are already determined.  Fixing the position of the
fourth vertex $v_i$ on plane $P_i$ causes the planarity constraints of
two inner quadrilaterals to combine to fix the position of $v_{i+1}$ on
plane $P_{i+1}$, etc: $v_{i+1} = f_i(v_i)$.  In order for the ring to be
formed correctly, these functions must return back to the starting
position: $v_0=f_3(f_2(f_1(f_0(v_0))))$.  But each $f_i$ is an affine
transformation, and we already know three noncollinear points of $P_0$
that are returned to themselves by the composition
$f_3(f_2(f_1(f_0(x))))$: the three original bicuboid vertices.  So the
composition must be the identity map, and {\em any} point of $P_0$ is
mapped by the functions $f_i$ to four points determining a geometrically
embedded ring.

The only remaining question is, when are these four points coplanar?
We say that a point $x$ in $P_0$ is {\em plane-generating}
when its quadruple $x$, $f_0(x)$, $f_1(f_0(x))$, $f_2(f_1(f_0(x)))$ forms a planar quadrilateral.
Then the plane-generating points form a linear subspace of $P_0$: if you have any two such
points $p$ and $q$, and a third point $r$ on line $pq$, then each point in the quadruple of $r$
is just the same linear combination of the images of $p$ and $q$,
so the quadruple of $r$ lies on a plane that is the same linear combination
of the planes for $p$ and $q$.  Note also that we already know two points
in $P_0$ that determine a coplanar quadrilateral: the vertices of the top
and bottom faces of the bicuboid.  So if there were a third affinely
independent point determining a coplanar quadrilateral then {\em all}
points of $P_0$ would determine a coplanar quadrilateral.
So the existence of a flat middle face for the $(2,0)$ mesh determines the
existence of a flat middle face for the $(0,2)$ hex and vice versa.
Since there are two degrees of freedom for placing the points of the
middle face, we can do so in such a way that the mesh remains generic.

Finally, consider the $(0,0)$--$(0,0)$ flip. The hexes of a $(0,0)$ mesh form a rhombic dodecahedron.
Color the degree-4 vertices of the 
dodecahedron gray, and color the degree-3 vertices alternately white 
and black, so that the cuboids in the given $(0,0)$ mesh have
one white vertex each.  To show that it can be 
subdivided into black cuboids, lift the vertices of the mesh into 
$\Real^4$ by assigning $w$-coordinates as follows.  Place the interior 
point $a$ (the apex) on the plane $w=1$ and the gray vertices 
$g_1,g_2,g_3,g_4,g_5,g_6$ on the plane $w=0$.  Each quadruple 
$ag_ig_jg_k$ determines a hyperplane, and these hyperplanes determine 
the $w$-coordinates of the black and white vertices.  The lifted 
cuboid mesh is the upper surface of a hypercuboid.  The single 
remaining hypercuboid vertex~$z$ (the zenith) is located at the 
intersection of four hyperplanes, each determined by a black point and 
its three gray neighbors.  Projecting the lower facets of the 
resulting hypercuboid back to $\Real^3$ gives us the flipped cuboid 
mesh.  In particular, the projection of $z$ lies in the interior of 
the original rhombic dodecahedron.
\end{proof}

The same argument shows that, for geometric hex meshes, regardless of
genericity, the $(3,0)$--$(0,3)$, $(0,3)$--$(3,0)$, $(2,0)$--$(0,2)$, $(0,2)$--$(2,0)$,
and $(0,0)$--$(0,0)$ flips are possible if and only if
the set of cuboids to be flipped forms a strictly convex subset of the
domain.

\SCfigure{unflippable}{Unflippable $(2,1)$ configuration.} {2in}

\SCfigure{unflip11}{Unflippable $(1,1)$ configuration.} {2in}

\SCfigure{quartz}{Vertex labels of $(0,1)$ configuration for proof of $(1,0)$ unflippability.}{2in}

However, not all flips are automatically flippable:

\begin{theorem}
\label{stickshift}
The $(1,1)$--$(1,1)$, $(2,1)$--$(1,2)$ and $(1,2)$--$(2,1)$ flips are not automatically flippable for geometric or self-intersecting meshes.  The $(1,0)$--$(0,1)$ flip is not automatically flippable for geometric meshes.
\end{theorem}

 \begin{proof}
One
can form a set of vertices in the pattern of a $(2,1)$ mesh
by placing four axis-aligned squares in four parallel planes: two
large ones above each other in the two inner planes, and two small ones
inside the projections of the large squares in the remaining planes
(Figure~\ref{unflippable}).  With this placement, the three-hex $(2,1)$
mesh has all faces flat.  However, the small squares can be
translated arbitrarily, causing the diagonal interior faces of the
five-hex $(1,2)$ mesh (such as the face formed by the four marked vertices in
the figure) to be warped, so that the $(2,1)$ mesh is not flippable.

Similarly, by centering these four squares on the same axis, but then shifting a pair of corresponding vertices in the two central squares along the edges connecting these two vertices to their counterparts in the outer two squares, one can create an unflippable $(1,2)$ mesh.

Next, choose parameters $r>0$ and $0<\theta<\pi/4$, and let $p(i,j)$ denote the point
$(r\cos i\theta,r\sin i\theta,j)\in\Real^3$.  Then the eight points
$p(\pm 1,0)$, $p(\pm 2,1)$, $p(\pm 3,1)$, $p(\pm 4,0)$ form the vertices
of a cuboid in which all facets are trapezoids (shown as the top eight vertices
of Figure~\ref{unflip11}).
If we make another copy of this set of eight vertices, translated downwards, and connect each bottom face of the cuboid to each bottom face of the copy, we get a $(1,1)$ mesh.
However, if the copies of $p(-2,1)$ and $p(-3,1)$ (the black marked points in the figure) are shifted less far than the copies of $p(2,1)$ and $p(3,1)$ (the grey marked points), then these four points will form a skew quadrilateral, preventing this mesh from being flipped.

Finally, consider the $(1,0)$ mesh, and label the $14$ vertices of the corresponding
$(0,1)$ mesh as illustrated in
Figure~\ref{quartz}.  Suppose also that the segments $p_iq_i$, $az$, and $by$ are all parallel.
We have the following equations for the two interior vertices $b$ and $y$:
\begin{eqnarray*}
b &= p_6p_1p_2 \cap p_2p_3p_4 \cap p_4p_5p_6 \\
y &= q_6q_1q_2 \cap q_2q_3q_4 \cap q_4q_5q_6
\end{eqnarray*}

Here, each triple denotes a plane through the three points.
Now consider shrinking the mesh by simultaneously contracting the parallel edges
$p_iq_i$ and  $az$.  As long as these edges have non-zero
length, the $(0,1)$ mesh is still valid.  But if we make them short
enough, $by$ is inverted, so the top and bottom cubes in the $(1,0)$ mesh
intersect. Thus, the $(0,1)$--$(1,0)$
flip is not automatically flippable.
\end{proof}

\begin{open}
Is the $(1,0)$ flip automatically flippable for self-intersecting meshes?  Is the $(0,1)$ flip automatically flippable?
\end{open}

Our results on automatic flippability are related to scene analysis
questions studied by Whiteley and Sugihara (see~\cite{Whi-DCG-89}). In
its most basic form, scene analysis asks whether a given planar drawing
of vertices connected by faces is the projection  of flat faces in
$\Real^3$. Here we are interested in whether a given set of hexes is the
projection of a cuboid in $\Real^4$, however our three-dimensional
scenes are not generic (due to the flatness constraints) so Whiteley's
formula for the number of degrees of freedom does not seem to apply. It
would be interesting to apply the  matroid techniques used in scene
analysis and rigidity theory to hex meshing.

\section{Bicuboid}\label{bicuboid-sec}

\SCfigure{bicuboid-fig}{The bicuboid.}{2.5in}

\OCfigure{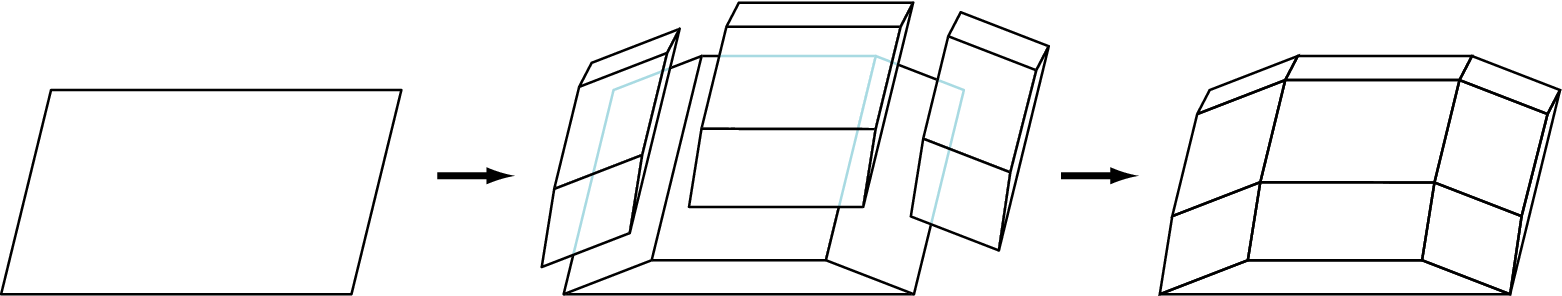}{3-by-3 subdivision of boundary can be
simulated by covering quads with four cuboids.}{5in}

\OCfigure{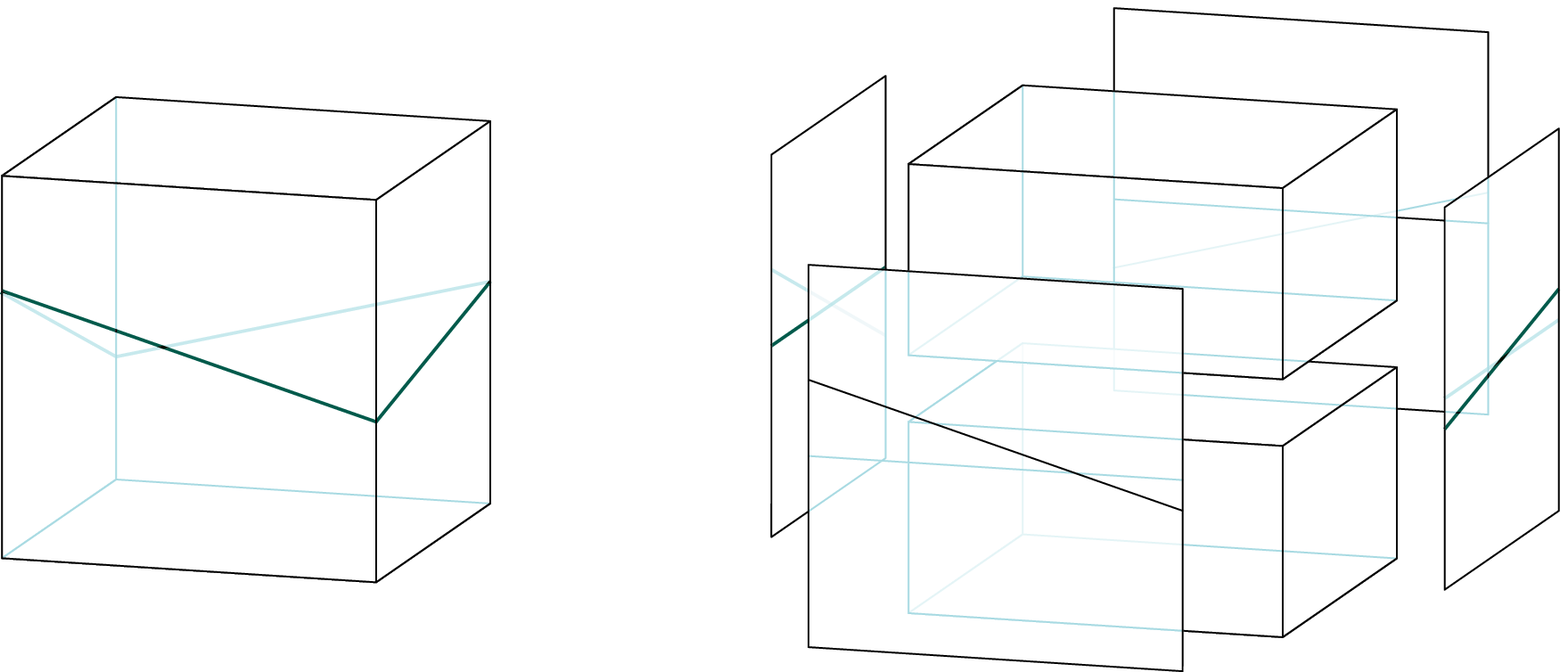}{Non-generic warped bicuboid (left)
with a self-intersecting mesh (right).}{4.5in}

In this section, we discuss a simple but important solid, called
a {\it bicuboid\/}. 
A bicuboid has the same topology as two hex elements sharing a face, as 
shown in Figure~\ref{bicuboid-fig}.  
We call a bicuboid {\it warped\/} if its central four vertices,
shown dotted in Figure~\ref{bicuboid-fig}, are not coplanar.
The bicuboid is important because, as we now show, the question of
whether topologically meshable domains have polyhedral hex meshes can be
reduced to this case, improving a reduction to a more complicated but
finite set of cases from our previous work~\cite{Epp-CGTA-99}.

\begin{theorem}\label{thm:bicuboid}
If every bicuboid has a geometric hex mesh, then any quad surface
mesh, with an even number of quad elements
and topologically equivalent to a sphere,
can be extended to a geometric hex volume mesh.
\end{theorem}

\begin{proof}
We start by finding a topological hex volume mesh
as in~\cite{Epp-CGTA-99,Mit-STACS-96,Thu-93}.
In such a mesh, each element is a topological cuboid,
meaning a cuboid with curved edges and faces.
We then subdivide all topological cuboids into 
$m \times m \times m$ grids of small topological cuboids
that are ``geometrically close'' to true cuboids. 
For example, if we require each face
of a small topological cuboid to fit within a thin slab
defined by parallel planes, we could pick a sufficiently large $m$
in order to make this possible.

Each small topological cuboid is then subdivided into seven pieces as
in a $(3,0)$--$(0,3)$ flip: a true 
cuboid in the center, and six ``almost-cuboidal'' warped hexes, each
with five planar faces and one nonplanar face.
The almost-cuboidal hexes of the subdivided mesh
match up in pairs to form bicuboids.  Meshing each bicuboid and
combining the results gives a mesh for the overall domain.

The only remaining problem is that the $m\times m\times m$ subdivision
also subdivides the boundary of the domain, which is not allowed.
To get around this problem, we let $m$ be a power of three,
and simulate the $3\times 3$ subdivision of a boundary quadrilateral by
placing four flat cuboids along it, as shown in Figure~\ref{3x3}, so that
the cuboids from adjacent boundary quadrilaterals meet face-to-face. The
number of degrees of freedom for the new cuboid vertices exceeds the
number of coplanarity constraints, allowing this construction to be
performed geometrically.
\end{proof}

Theorem~\ref{thm:bicuboid} shows that a bicuboid is
the crucial test case for the question of whether every
geometric quad surface mesh can be extended to a geometric hex mesh.
Theorem~\ref{automatic}, however, leads us to suspect that 
the answer is no, since it shows that a warped bicuboid cannot be
meshed by any generic mesh that can be reached from the two-hex mesh by

automatically flippable flips---reversing those flips would give a
geometric two-hex mesh, which does not exist.

\begin{open}
Does every warped bicuboid have a geometric hex mesh?
Does every warped cuboid have a generic
self-intersecting hex mesh?
\end{open}

The question is not open for non-generic self-intersecting meshes: a
warped cuboid formed by placing the four vertices of a warped
quadrilateral around the equator of a unit cube has a non-generic
self-intersecting six-hex mesh in which the inner quadrilateral's
vertices also lie on the surface of the cube (Figure~\ref{ngwbc}).
Less degenerately, if one has a warped bicuboid such that the convex hull of its top and bottom quadrilaterals forms a cuboid, then the toroidal difference between the bicuboid and this central cuboid can be partitioned into four triangular prisms; if we then slice the central cuboid in two by a plane, this plane divides a face of each prism in two, forming degenerate cuboids each with a single $180^\circ$ dihedral.

\section{Parity}
\label{parsec}

\OCfigure{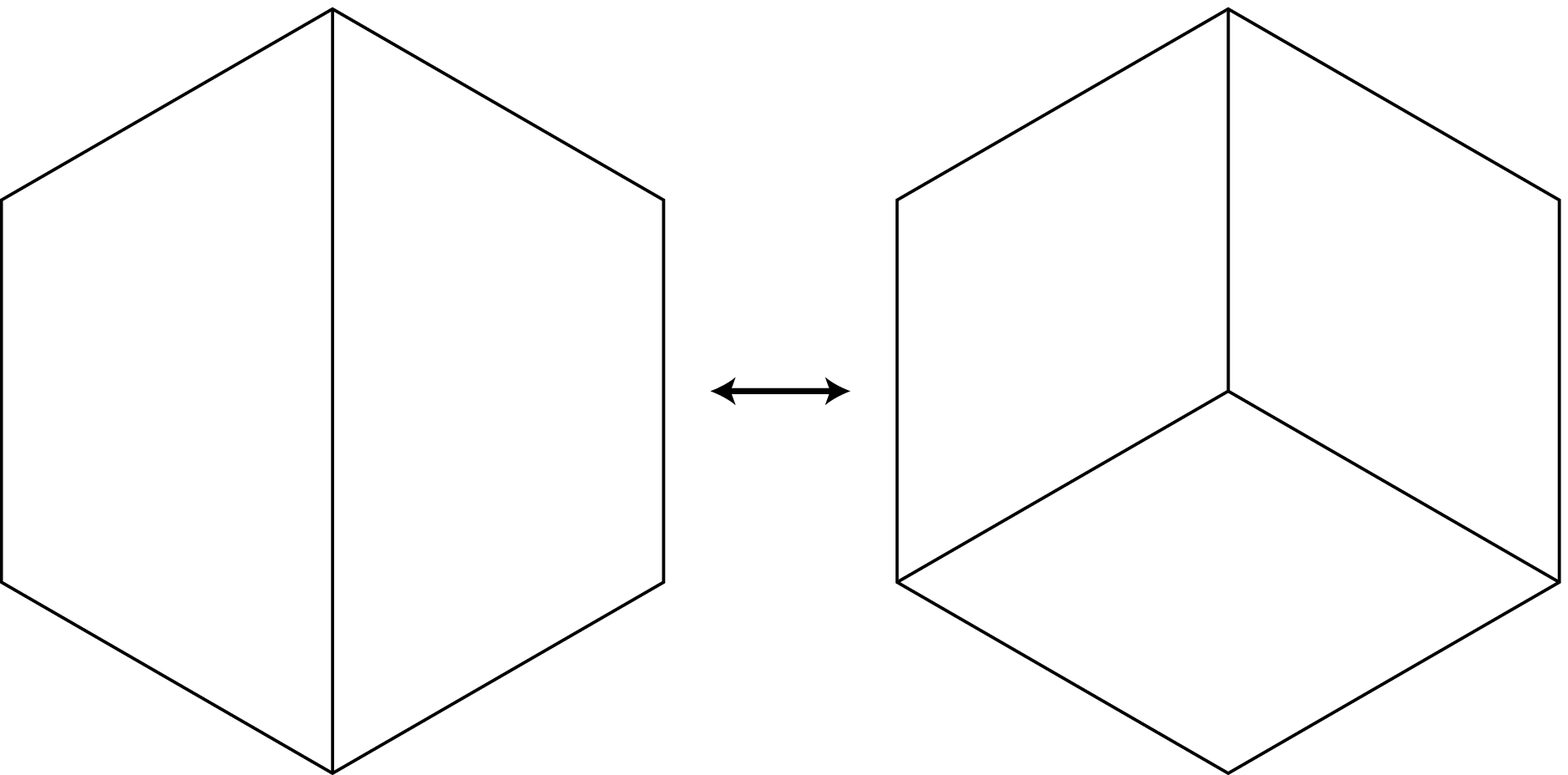}{Parity-changing transformation of a quadrilateral
mesh.}
{3.25in}

\OCfigure{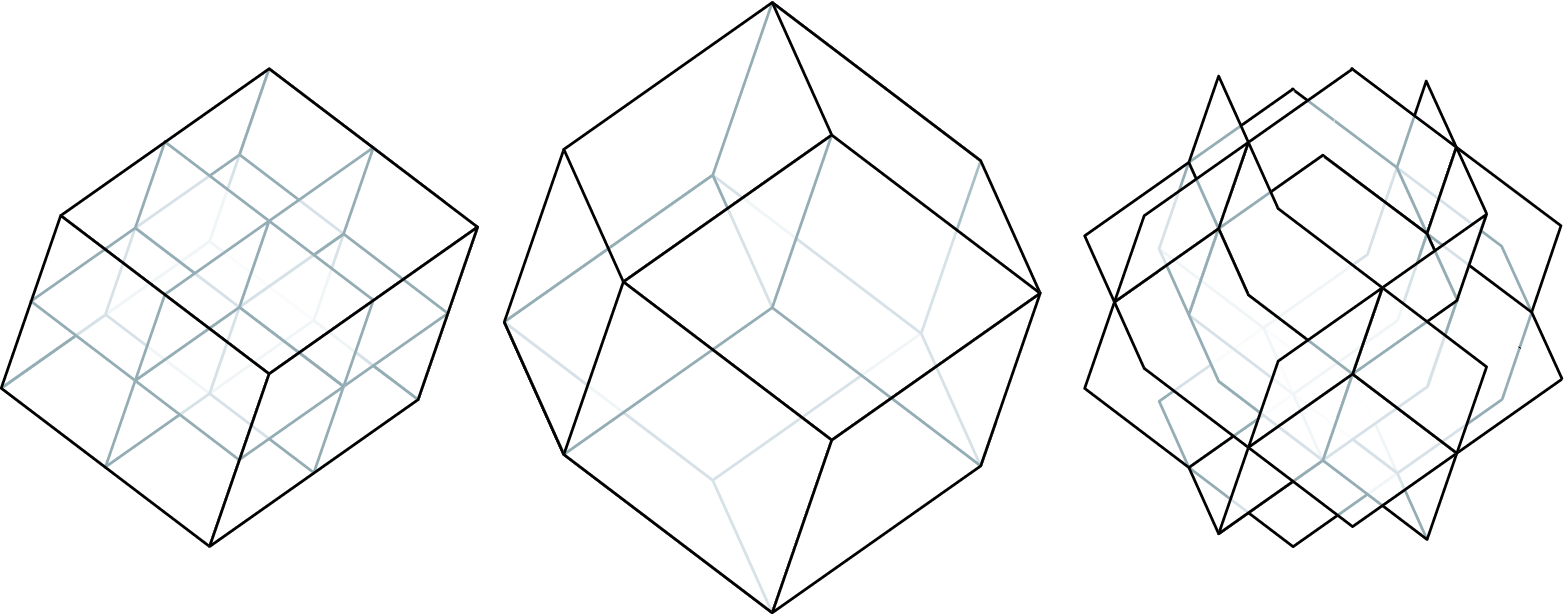}{Cuboid divided into octants by three dual surfaces
(left), and the set of dual surfaces (right) for a more complicated
four-hex mesh (center).}{5in}

Each of the flip transformations preserves
the parity (odd or even) of the number of hex elements in the mesh.
In two dimensions, any meshable domain has meshes of both parities, as
can easily be seen by performing a two-to-three quad replacement (not a
flip), as shown in Figure~\ref{quad2to3}, so not all quad
meshes are obtainable by flipping from a given starting mesh.
A similar result holds for hex meshes, but the construction is more
complicated: 

\begin{theorem}\label{parity}
If a given quad surface mesh can be extended to a topological hex mesh,
it has topological hex meshes with both even and odd numbers of
hexes. 
\end{theorem}

\begin{proof}
Suppose we have some hex mesh, and we wish to find an alternative mesh
with the opposite parity.  Following Mitchell~\cite{Mit-STACS-96} and
Thurston~\cite{Thu-93}, we form three quadrilateral surface patches
within each hex, medial between each of the three pairs of opposite
faces; these patches join up to form an arrangement of surfaces, with
one triple intersection per cuboid (Figure~\ref{hexdual}).  The
boundaries of these surfaces form a graph on the domain boundary that is
the dual of the original quad surface mesh.

Conversely, if we have an arrangement of surfaces that meets the domain
boundary in the dual of the quad mesh, where at most three surfaces meet
in any point and we have no pinch points or other topological anomalies,
we can form a graph with one vertex per triple intersection and with
edges connecting the vertices along the curves where two surfaces meet. 
As long as this graph is connected and has no self-loops or multiple
adjacencies, we can form a valid topological hex mesh whose cuboids and
quadrilaterals correspond to the vertices and edges of this
graph.  If the graph does have self-loops or multiple adjacencies,
Mitchell~\cite{Mit-STACS-96} shows that they can be removed by the
``pillowing'' ($(3,0)$--$(0,3)$ flip) operation.

In order to change the parity of the number of hexes in the mesh, then,
we can introduce a new surface in such a way that the number of triple
intersections increases by an odd number.  A suitable candidate is Boy's
surface~\cite{Boy-MA-03,Fra-87}, a projective plane embedded in
$\Real^3$ in such a way that there is one triple intersection point,
connected to itself by double-intersection curves in a pattern of three
self-loops.  We place a copy of Boy's surface within a
simply-connected region of the dual arrangement, in such a way that the
augmented arrangement remains connected and continues to avoid multiple
adjacencies.  Then each of the three loops of Boy's surface will cross
the other arrangement surfaces an even number of times, and the
double intersection curves of those other surfaces will cross the new
Boy's surface an even number of times, so that the only change in parity
will be from the single triple intersection of Boy's surface with
itself.  The resulting graph may have some self-loops or multiple
adjacencies, but we can remove these by pillowing without further
changes of parity.  The hex mesh formed by this graph will therefore
have the opposite parity from the mesh we started with.
\end{proof}

\SCfigure{Boy}{Boy's surface.}{2.5in}

\begin{open}
Are there domains with geometric hex meshes of both parities?
\end{open}

For quadrilateral meshes, one can allow parity-changing operations by
adding Figure~\ref{quad2to3} to the repertoire of flips.

\begin{open}
Is there a simpler parity-changing operation for hexahedral meshes?
What is the minimum possible number of hexahedra involved in such an operation?
\end{open}

For non-simply-connected domains, parity changing may be easier: for instance, a torus with a hexagonal cross section and an odd number of segments may be filled with either two or three hexes per segment, as in the planar parity-changing operation.  For instance, with three segments, this leads to an even mesh with six hexes and an odd mesh with nine.  However such a torus cannot be part of a mesh of a simply-connected domain because the complement of the torus would have a contractible odd cycle.

\section{Bubble-wrapping and Shelling}

\SCfigure{quaddual}{Curve arrangement dual to a quadrilateral
mesh.}{3.5in}

\OCfigure{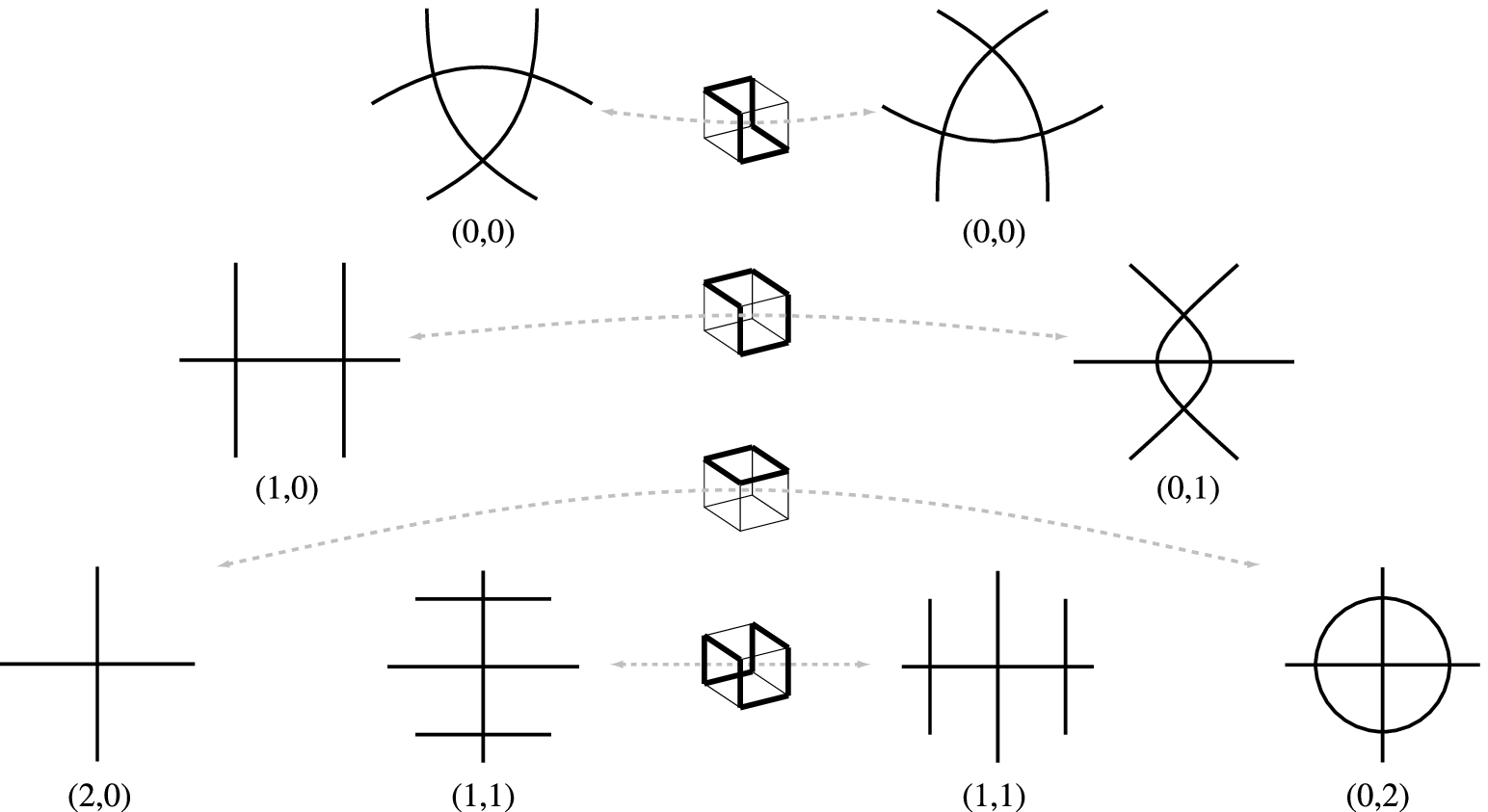}{Actions of quadrilateral flips on dual curves.}{6.5in}

In order to prove our results in the next section on flip graph
connectivity, we need a technical strengthening of the Mitchell-Thurston
result on topological mesh existence: every ball-shaped domain with an
even number of quadrilateral faces has a {\em shellable} hexahedral mesh,
where a {\em shelling} of a complex~\cite{Zie-95} is an ordering of its
cells such that the union of the cells in any suffix of the ordering
forms a simply-connected set, and such that each cell meets the union of
the set of cells coming after it in the ordering in a simply-connected
subset.  In this section we prove a simpler version of this
strengthening: define a {\em pseudo-shelling} to be an ordering of the
cells such that the union of the cells in any suffix of the ordering is
simply connected, without the requirement on how each cell meets this
union.  That is, we want to remove elements one by one from our mesh in
such a way that the remaining elements always form a ball-shaped
domain.  Then we prove here that every simply-connected domain with an
even number of quadrilateral faces has a pseudo-shelling.

To do this, we take a dual view similar to that in
Theorem~\ref{parity} and in~\cite{Mue-EC-99}. For any topological quadrilateral
mesh on a topological-sphere surface, define a curve arrangement by
connecting each opposite pair of edge midpoints in each quadrilateral by
a segment of a curve (Figure~\ref{quaddual}).  This is a {\em simple}
arrangement, meaning that at most two curves intersect at any point, and
whenever two curves intersect they cross each other at a single point.
Because the surface has no boundary, the curves have no endpoints; each
curve crossing corresponds to a mesh element.  Each element removal in a
pseudo-shelling corresponds to a flip in the boundary quadrilateral
mesh, and we consider these flips in terms of their effect on the dual
curve arrangement (Figure~\ref{dualjqf}).  By finding a sequence of flips that can be used to
transform any such arrangement into the arrangement dual to a single
cube's boundary, we will show that a pseudo-shelling exists.

\SCfigure{bubblewrap}{Bubble-wrapping a curve arrangement}{4in}

\begin{lemma}
A simple curve arrangement is dual to a quadrilateral mesh if and only if
the graph formed by its crossing vertices and edges is 3-connected.
\end{lemma}

\begin{proof}
Any simple curve arrangement can be dualized by forming a quadrilateral
for each crossing, however the resulting collection of cells must meet
face-to-face and vertex-to-vertex to be a cell complex.
 It is not hard to see that, if the graph has
one or two vertices that separate it into nontrivial components, then
it must have a degeneracy that prevents it from being a cell complex;
for instance, a single articulation point corresponds to a
quadrilateral that is adjacent to itself along a pair of edges.
Conversely, if an arrangement forms a 3-connected planar graph, then by
Steinitz' theorem~\cite{Zie-95} it can be represented as the vertices and edges of a
convex polyhedron, and the polar polyhedron's facets form the desired
dual mesh.
\end{proof}

Since not all arrangements are 3-connected, it is convenient to work
with a modification of the mesh that is guaranteed to be 3-connected.
For a given arrangement $A$, we define the {\em bubble-wrapped
arrangement} $B(A)$ by adding curves to $A$ as follows: first, form a
small circular curve around each vertex of $A$.  Second, repeat this
step around each new crossing point formed by the new circles and the
edges of $A$.  Each edge has two circles added in this second step, and
these circles should be drawn so that they cross each other, and so that
the edge passes through their intersection.  Figure~\ref{bubblewrap}
depicts these bubble-wrapping steps.

\SCfigure{arrops}{Three operations on arrangement: pulling apart or
pushing together (left); inverting a triangle (middle), and switching
(right).}{3.5in}

\begin{lemma}
If $A$ is a connected arrangement, $B(A)$ is 3-connected.
\end{lemma}

\begin{lemma}\label{wrappable}
If $A$ is the dual to a quadrilateral mesh, $B(A)$ can be formed by
flips from $A$.
\end{lemma}

\begin{proof}
We create all the circles used in the bubble-wrapping process by $(2,0)$--$(0,2)$
flips, and cross the two circles on each edge by a $(1,0)$--$(0,1)$ flip.
It is not hard to see that each step in this process preserves
3-connectivity of the arrangement.
\end{proof}

We now define three operations that modify the topology of a
curve arrangement in simpler ways than the modifications performed by the flip duals; in some ways these operations are similar to the well-known Reidemeister moves for knots in $\Real^3$, or the Whitney moves for curves in $\Real^2$ \cite{MehYap-SJC-91,Whi-CM-37}.  If the arrangement has a pair of crossings that are
connected by two edges, without other curves between them, we can {\em
pull apart} the crossings and form an arrangement with two fewer
vertices; conversely we can {\em push together} two nearby curves to add
two vertices to an arrangement (Figure~\ref{arrops}, left).
If the arrangement has a face formed by a triangle of three crossings
and three curves, we can {\em invert} the triangle by passing one curve
across the crossing of the other two (Figure~\ref{arrops}, middle).
Finally, if the arrangement contains two nearby connected and
non-crossing segments of curves, we can {\em switch} these segments by
removing them from the arrangement and reconnecting the curves by two
other non-crossing segments (Figure~\ref{arrops}, right).

\begin{lemma}\label{wrapped-ops}
Suppose two connected arrangements $A$ and $A'$ are related to each
other by one of the three operations described above.  Then $B(A)$ can
be transformed into $B(A')$ by a sequence of flips.
\end{lemma}

\SCfigure{invbub}{Steps in inversion of a triangle in a bubble-wrapped
arrangement.}{4.5in}

\begin{proof}
We begin by describing the simplest case, inversion of a triangle.
If $A$ were not bubble-wrapped, we could perform this operation by a
single $(0,0)$--$(0,0)$ flip, however the difficulty is in rearranging the
bubble-wrap curves to allow this flip without violating 3-connectivity.
Suppose
$A$ contains a triangle that can be inverted; its bubble-wrapped version
$B(A)$ is depicted in Figure~\ref{invbub} (left). We perform a $(2,0)$--$(0,2)$ flip
at one of the crossings of $B(A)$ in this triangle, adding a new curve
to the arrangement, and perform a sequence of $(1,0)$--$(0,1)$ and $(0,0)$--$(0,0)$ flips to
extend the region interior to this new curve until it covers the whole
triangle without extending outside the bubble-wrapping of the triangle
(Figure~\ref{invbub}, middle).  Next, we remove the nine curves forming
the bubble-wrapping around the triangle, by using $(0,0)$--$(0,0)$ and $(0,1)$--$(1,0)$ flips
to reduce the size of the region each one contains until it can be
removed by a $(0,2)$--$(2,0)$ flip (Figure~\ref{invbub}, right).  At this point the
triangle is exposed in the interior of the newly added curve and can be
inverted by a $(0,0)$--$(0,0)$ flip, after which we reverse the above steps to
restore the bubble-wrapping around the triangle.  It is not difficult to
verify that each step in this process can be performed while preserving
3-connectivity of the overall arrangement.

Similarly, pulling apart and pushing together can be performed by
modifying the bubble-wrapping to remove the eight curves protecting the crossings and edges in the pulled-apart region, and replace them by a single curve surrounding an adjacent face of the arrangement,
in a way that allows the change to $A$ to
be made by a single $(1,0)$--$(0,1)$ or $(0,1)$--$(1,0)$ flip.
Switching can be performed by adding curves around the face of the arrangement between the switched edges, so that one curve crosses both switched edges and another curve passes between them, in a way that allows the change to $A$
to be made by a single $(1,1)$--$(1,1)$ flip.  The details are more complicated
than inversion, and we omit them.
\end{proof}

\SCfigure{switchcross}{Switching near a crossing.}{2.5in}

\SCfigure{switchloop}{Switching near a loop.}{3.75in}

\begin{lemma}
Any simply-connected domain, with a boundary consisting of an even
number of quadrilaterals, has a mesh with a pseudo-shelling.
\end{lemma}

\begin{proof}
As stated at the start of the section, we describe the pseudo-shelling
in terms of the effect each removal of a hexahedral element
causes to the curve arrangement $A$ dual to the domain boundary. By
Lemma~\ref{wrappable}, we can form an initial sequence of flips to
bubble-wrap $A$.  Next, we apply Lemma~\ref{wrapped-ops} to perform a
sequence of switching operations near each crossing of $A$.  Each such
operation creates a loop connecting the crossing to itself (Figure~\ref{switchcross}), and reduces
the number of crossings without such loops; it is always possible to
choose a way to perform the switching operation that preserves the
connectivity of the overall arrangement.

\SCfigure{loopcancel}{Removing two adjacent loops by a push,
invert, and two pull operations (the Whitney trick).}{5in}

After this step, the transformed arrangement must consist of a single
simple closed curve, decorated with a sequence of loops.  Since $A$
initially had an even number of crossings, and each flip preserves
parity, there are an even number of loops along the curve.  Note that
a switching operation near the crossing of a loop can result in a new loop
on the other side of the simple closed curve (Figure~\ref{switchloop}).  We perform additional
switching operations near the loops' crossings if necessary so that the
loops can be grouped into adjacent pairs, with one member of each pair
interior to the simple closed curve and one member exterior.  Each pair
can be removed by a sequence of a push, triangle inversion, and two pull
operations (Figure~\ref{loopcancel}); this sequence of operations is known to topologists as the {\em Whitney trick}.  After repeated Whitney tricks, the arrangement is transformed into a simple closed curve protected by bubbles; we remove bubbles until only two are left, forming
the dual arrangement of a cube, the desired final state of a
pseudo-shelling.
\end{proof}

The existence of a mesh with a shelling, rather than just a
pseudo-shelling, can be deduced via the methods used to prove
Theorem~\ref{thm:qfc} below.

\section{Connectivity}

The results of Section~\ref{parsec} naturally raise the question
whether parity is the only obstacle to reachability by flips, or whether
there might be some more subtle property of a mesh that prevents it being
formed by flipping from some other mesh.  Phrased another way, can every
mesh transformation that replaces some bounded submesh by another be
simulated by a sequence of flips?  Figure~\ref{hardtoflip} depicts
a difficult example, closely related to a hexahedral meshing problem
posed by Schneiders~\cite{Sch-Open}.  We leave finding a flip sequence
for this example as a puzzle for the reader.

For any domain with boundary mesh, and a type of mesh to use for
that domain, define the {\em flip graph}~\cite{San-JAMS-00} to be a graph with (infinitely
many) vertices corresponding to possible meshes of the
domain, and an edge connecting two vertices whenever the corresponding
two meshes can be transformed into each other by a single flip.  In this
framework, the question above can be phrased as asking for a description
of the connected components of the flip graph.

\begin{theorem}\label{thm:qfc}
The flip graph for topological quad meshes of any simply-connected
domain has exactly two connected components.
\end{theorem}

\begin{proof}
Due to parity, there must be at least two components. It remains to show
that any two meshes $M_1$ and $M_2$ of the same parity can be flipped
into each other. Consider a three-dimensional ball, the upper hemisphere
of which is partitioned into mesh $M_1$ and the lower hemisphere of
which is partitioned into mesh $M_2$, the two meshes meeting along the
equator of the ball at corresponding boundary edges.  Because $M_1$ and
$M_2$ have the same parity, the ball has an even number of quadrilateral
faces and so has a topological hex mesh $H$.

The possible existence of
pairs of quadrilaterals meeting in more than one edge (along the ball's
equator) does not lead to difficulties with degenerate meshes,
as we can remove these degeneracies by performing $(2,0)$--$(0,2)$ and $(0,2)$--$(2,0)$ flips in
$M_1$ and $M_2$ respectively.

We now wish to remove the cuboids from $H$ one by one,
reducing the initial ball to successively smaller shapes.  At each step
in the removal process, we will maintain a mesh $M$ describing 
describing the upper boundary of the remaining shape; initially
$M=M_1$.  As long as each removed cuboid touches the upper boundary of $M$ in a simply
connected set of faces, the change to $M$ caused by its removal is a
flip.  If we can remove all the cuboids in this way, we will have
flipped $M_1$ into $M_2$.

The requirement of simply-connected incidence with $M$ for each removed
cuboid is very similar to the definition of a {\em shelling} of a
complex, as described in the previous section; it differs from a shelling only in the requirement that cuboids be removed from the upper boundary.
As shown in the previous section, we can at least choose $H$ to
be pseudo-shellable.  We then partition the pseudo-shelling sequence into
{\em layers} of independent sets of hexahedral elements, such that no two elements in the
same layer share a facet.  Let $B_i$ denote the ball formed by the union
of layer $i$ and all successive layers. The boundary of each $B_i$ is a
quadrilateral mesh; we thicken each quadrilateral of this mesh into a
hexahedron, forming another layer of hexahedra separating layer $i$ from
layer $i-1$.  In terms of the dual surface arrangement this corresponds
to adding a topological-sphere surface surrounding $B_i$.
After performing this thickening process, we have a mesh with
alternating nested layers of cells: half of the layers consist of cells
connecting two nested and combinatorially equivalent quad meshes, while
the other half of the layers consist of cells not sharing any facets
with each other.

We order the cells of this mesh as follows: start by choosing some cell
adjacent to
$M_1$, then some cell in the next layer adjacent to the starting cell,
and continue ``drilling down'' by removing at most one cell from each
layer until reaching the center of the set of concentric spheres.  Then,
remove the layers from the inside out, removing all cells from a given
layer before starting the next outer layer.  The existence of an
appropriate ordering within each layer follows from the existence of
shellings for all planar complexes.
\end{proof}

\SCfigure{hardtoflip}{Transformation of one four-quad
mesh into another. Can it be simulated by a sequence of flips?}{3in}

\OCfigure{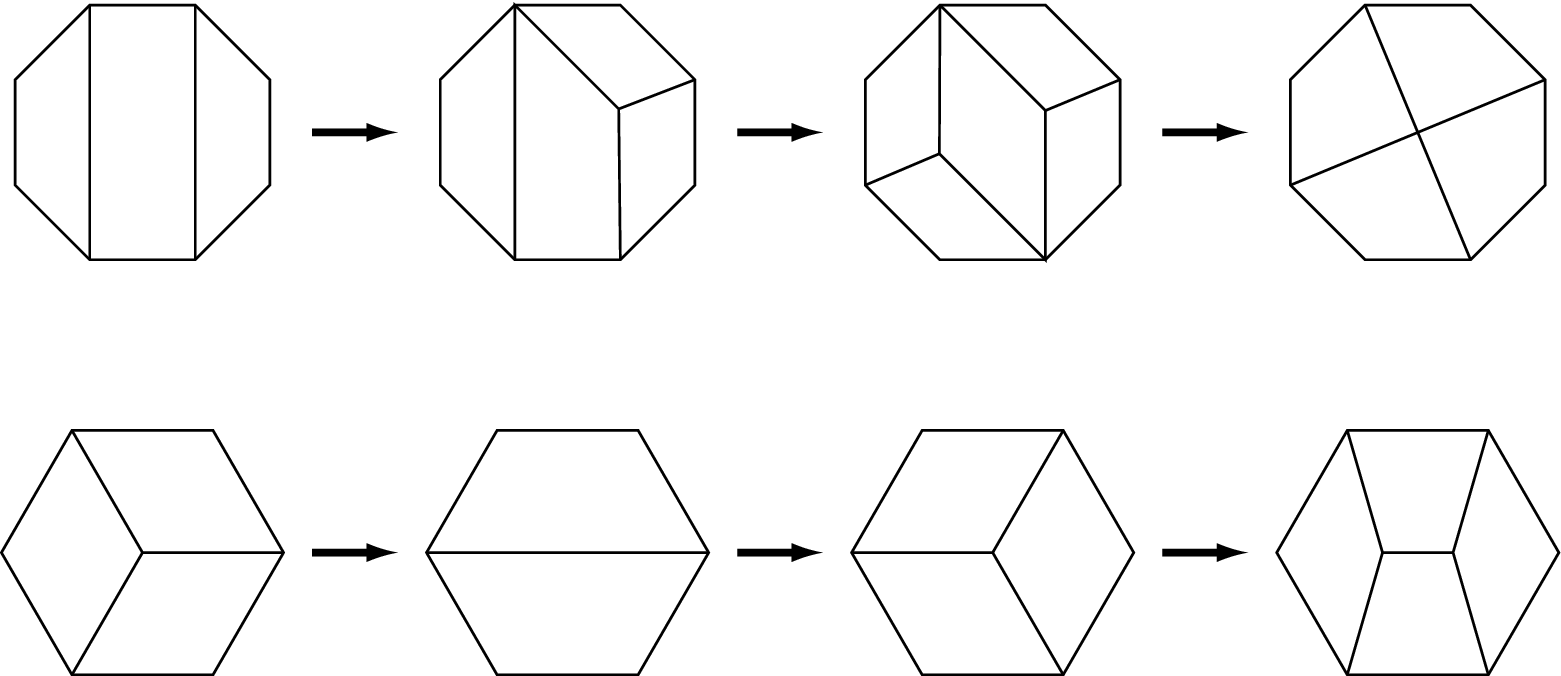}{Simulation of flips by parity-change operation:
half of $(1,1)$--$(1,1)$ flip (top), $(0,0)$--$(0,0)$ flip (bottom, first and third positions), and
$(1,0)$--$(0,1)$ flip (bottom,
second and fourth positions).}{5.25in}

\OCfigure{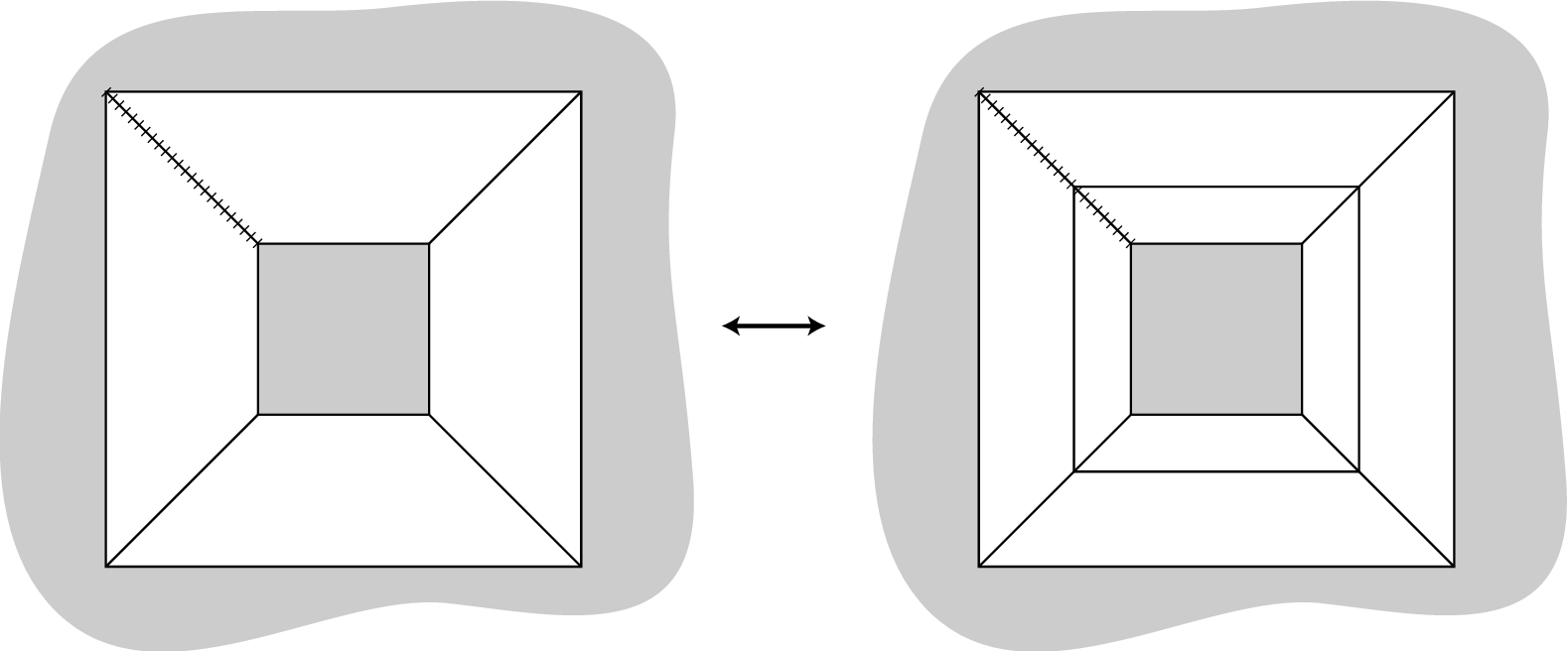}{Transformation of a non-simply-connected quad mesh
that cannot be achieved by flips. The three edges indicated by hash
marks form a contractible triangle in the region
bounded above and below by the left and right meshes.}{5in}

As a consequence, the set of flips together with the parity-changing
operation depicted in Figure~\ref{quad2to3} form a complete set of local
mesh transformations, in that they are able to simulate any other
transformation operation.  In fact, since the parity-changing operation
can simulate many of the possible flips (Figure~\ref{simflips}),
it and the $(2,0)$--$(0,2)$ flip together form a complete set.

Combining the proof method with our previous linear-complexity
hex meshing techniques~\cite{Epp-CGTA-99} shows that any two equal-parity
quad meshes $M_1$ and $M_2$ of a simply-connected region can be flipped
one to the other (as topological meshes) in a number of steps linear in
the total number of cells in the two meshes.

Conversely to the techniques of this proof, any sequence of flips from
one quad mesh to another can be seen as forming a hex mesh of a three
dimensional region bounded above and below by the two meshes. The
assumption of simple connectivity is necessary for the result, as can be
seen from Figure~\ref{unflipq}: the region bounded by the two
meshes has odd cycles on its boundary (triangles formed by
one diagonal edge of the four-quad mesh and two edges of the eight-quad
mesh) that can be contracted to points in the interior of the region;
Mitchell~\cite{Mit-STACS-96} has shown that such contractible odd cycles
make hex meshing an impossibility. Therefore there is no way to
flip one mesh to the other, despite their having equal
parity.

\begin{open}
Can we characterize the flip graphs for geometric quad meshes of non-simply-connected regions?  Is the number of components of the flip graph determined by the number of topological holes in the region?
\end{open}

For the analogous problem of geometric triangle mesh flip graph
connectivity, a single connected component is known to exist due to the
ability to flip to the Delaunay triangulation, and the ability to
add or remove vertices by changing the weights of a regular
triangulation.  However, higher dimensional flip graphs can be disconnected~\cite{San-JAMS-00}.  There has also been some work on finding efficient
sequences of
flips~\cite{HanOttSch-JUCS-96,HurNoyUrr-DCG-99,SleTarThu-JAMS-88}
although finding the shortest such sequence even for triangulations of
convex polygons remains a major open problem.

\begin{open}
Can the flip graph for topological or geometric hex meshes of a simply
connected region have more than two components?
\end{open}

Answering this question seems to require a four dimensional
generalization of Mitchell and Thurston's result.  Even the tetrahedral
mesh version of this problem seems difficult: topological tetrahedron
mesh flip graphs are connected~\cite{Lic-KF-99},
and flipping can be used to form Delaunay
triangulations~\cite{EdeSha-Algo-96,Joe-CAGD-91}, but this does not seem
to imply geometric tetrahedral flip graph connectivity since the
flipping-based Delaunay construction algorithms
do not allow the initial triangulation or sequence of flips to be chosen
arbitrarily.

\section{Conclusions}\label{disc-sec}

In this paper, we have introduced (or at least popularized)
flipping moves for quad and hex meshes.  We have also considered the
conditions under which flips are realizable, tested the connectivity of
the flip graph, and used flipping to show some evidence that not all quad
surface meshes can be extended to  hex volume meshes. 

\section*{Acknowledgements}

A preliminary version of this paper appeared at the 10th International
Meshing Roundtable.  Work of Eppstein was supported in part by NSF grants
CCR-9258355 and CCR-9912338
and by matching funds from Xerox Corp., and performed in
part while at Xerox PARC.
Work of Erickson was supported in part by a Sloan Fellowship and by NSF grants CCR-0093348 and DMR-0121695.
We thank Scott Mitchell, Scott Canann, and the
Roundtable referees for many helpful comments.

\raggedright
\bibliographystyle{abuser}
\bibliography{hexmesh}

\end{document}